\documentclass[12pt, draftclsnofoot, onecolumn]{IEEEtran}

\ifCLASSINFOpdf
  \usepackage[pdftex]{graphicx}
  \DeclareGraphicsExtensions{.pdf,.jpeg,.png}
\else
  \usepackage[dvips]{graphicx}
  \DeclareGraphicsExtensions{.eps}
\fi

\usepackage{dsfont}

\usepackage[fleqn]{amsmath}
\usepackage{amssymb}
\usepackage{empheq}
\usepackage{color}
\usepackage{algorithmic}
\usepackage[]{algorithm}
\usepackage{enumerate}
\usepackage{relsize}

\renewcommand{\refeq}[1]{(\ref{#1})}

\newtheorem{remark}{Remark}

\newtheorem{definition}{Definition}
\newtheorem{fact}{Fact}
\newtheorem{assumption}{Assumption}

\newcommand{\field}[1]{\mathbb{#1}}
\newcommand{\ve}[1]{{\mathbf{#1}}}
\newcommand{\ma}[1]{{\mathbf{#1}}}
\newcommand{\set}[1]{{\mathcal{#1}}}
\newcommand{\NN}{\field{N}}

\newcommand{\RN}{{\field{R}}_+}

\newcommand{\R}{{\field{R}}}

\newcommand{\sets}{{\set{S}}}

\newcommand{\setm}{{\set{M}}}

\newcommand{\setn}{{\set{N}}}
\newcommand{\setl}{{\set{L}}}

\newcommand{\BEAS}{\begin{eqnarray*}}
\newcommand{\EEAS}{\end{eqnarray*}}
\newcommand{\BEQ}{\begin{equation}}
\newcommand{\EEQ}{\end{equation}}
\newcommand{\BIT}{\begin{itemize}}
\newcommand{\EIT}{\end{itemize}}
\newcommand{\BAR}{\begin{array}}
\newcommand{\EAR}{\end{array}}

\newcommand{\argmax}{\mathop{\rm argmax}}

\newcommand{\signal}[1]{{\boldsymbol{#1}}}

\newcommand{\Natural}{{\mathbb N}}

\newif\ifcomments
\commentsfalse

\usepackage{todonotes}
\usepackage[color]{changebar}
\usepackage[normalem]{ulem}
\cbcolor{red}

\usepackage{caption}
\usepackage{subcaption}

\renewcommand{\baselinestretch}{1.417}

\ifcomments
\newcommand{\removed}[1]{\cbstart\removedfragile{#1}\cbend{}}
\newcommand{\removedfragile}[1]{{\color{red}{\sout{#1}}}{}}
\newcommand{\added}[1]{\cbstart\addedfragile{#1}\cbend{}}

\newcommand{\changed}[2]{\added{#1}\removed{#2}}

\newcommand{\quest}[1]{\todo[author=General Question,inline,color=yellow!40,caption={}]{#1}}
\newcommand{\todoSL}[1]{\todo[author=Task
  S\l{}awomir,inline,color=red!70,caption={}]{#1}}
\newcommand{\todoEM}[1]{\todo[author=Task
  Emmanuel,inline,color=green!40,caption={}]{#1}}
\newcommand{\todoRE}[1]{\todo[author=Task Renato,inline,color=blue!40,caption={}]{#1}}

\newcommand{\note}[1]{\todo[author=Note,inline,color=blue!40,caption={}]{#1}}
\else
\newcommand{\added}[1]{#1}
\newcommand{\removed}[1]{}
\newcommand{\changed}[2]{\added{#1}}
\newcommand{\cosl}[1]{}
\newcommand{\note}[1]{}
\newcommand{\quest}[1]{}
\newcommand{\todoSL}[1]{}
\newcommand{\todoEM}[1]{}
\newcommand{\todoRE}[1]{}
\fi

\newcounter{assume}

\begin{document}
\title{Traffic Demand-Aware Topology Control for Enhanced
  Energy-Efficiency of Cellular Networks}

\author{Emmanuel~Pollakis,~\IEEEmembership{no Member,~IEEE,}
        Renato~L.~G.~Cavalcante,~\IEEEmembership{Member,~IEEE,}
        and~S\l awomir~Sta\'nczak,~\IEEEmembership{Senior~Member,~IEEE}
\thanks{Emmanuel~Pollakis, Renato~L.~G.~Cavalcante and Slawomir~Stanczak are with the Fraunhofer Institute for Telecommunications,
Heinrich Hertz Institute, 10587 Berlin, Germany (e-mail: \{emmanuel.pollakis, renato.cavalcante, slawomir.stanczak\}@
hhi.fraunhofer.de).}

\thanks{Parts of the material in this paper were presented at the 2012
  IEEE Signal Processing Advances Wireless Communications (SPAWC)
  Workshop, Cesme, Turkey, 17-20 June 2012 \cite{Pollakis2012}
	.}

}

\maketitle

\begin{abstract}
  The service provided by mobile networks operated today is not
  adapted to spatio-temporal fluctuations in traffic demand, although
  such fluctuations offer opportunities for energy savings. In
  particular, significant gains in energy efficiency are realizable by
  disengaging temporarily redundant hardware components of base
  stations. We therefore propose a novel optimization framework that
  considers both the load-dependent energy radiated by the antennas
  and the remaining forms of energy needed for operating the base
  stations. The objective is to reduce the energy consumption of
  mobile networks, while ensuring that the data rate requirements of
  the users are met throughout the coverage area. Building upon sparse
  optimization techniques, we develop a majorization-minimization
  algorithm with the ability to identify energy-efficient network
  configurations. The iterative algorithm is load-aware, has low
  computational complexity, and can be implemented in an online
  fashion to exploit load fluctuations on a short time
  scale. Simulations show that the algorithm can find network
  configurations with the energy consumption similar to that obtained
  with global optimization tools, which cannot be applied to real
  large networks.  Although we consider only one currently deployed
  cellular technology, the optimization framework is general,
  potentially applicable to a large class of access technologies.
\end{abstract}


\IEEEpeerreviewmaketitle

\section{Introduction}
\label{sec:intro}  

\IEEEPARstart{M}{obile} networks are constantly growing, 
with the number of mobile subscribers worldwide exceeding 7.1 billion in
2014 and the current forecast of 9.5 billion subscribers in 2020
\cite{Ericsson2014}. During this time, an even more drastic
increase is predicted for mobile data traffic, which is
  expected to grow by the factor of 8. To keep pace with this
  trend, mobile network operators attempt to boost the network capacity 
	by deploying additional antennas and base stations. This will
lead to a dramatic increase in the energy consumption, thereby 
causing additional operational expenses (OPEX) and higher
carbon dioxide (CO$_2$) emissions. Therefore, enhancing the energy
efficiency of mobile communications systems has a positive double 
effect: lower greenhouse gas emissions and reduced OPEX.

Current mobile networks have been designed
to provide the best possible service to the
  users at all times. Global network parameters and the network
topology are largely static, although, as pointed out in many studies (see for
instance \cite{Willkomm2009, Corliano2008, Oh2011,
  Alcatel-Lucent2008}), the traffic load fluctuates
significantly over time and space. Such spatio-temporal
  fluctuations create large capacity surpluses at times
of low traffic demand, which in turn offers opportunities for energy
savings through adaptation of the service supply to the actual demand. 
However, in order to utilize the capacity surpluses for significant 
energy savings, it is essential to reduce the energy consumed by hardware and
auxiliary equipment (e.g. coolers), which is a dominant
form of energy consumption in current mobile networks. In fact, 
for a typical network with today's technology, base stations 
consume over 50\% of the total network energy budget \cite{Han2011}. 
Most of the energy consumed by base stations is spent on powering hardware and
  auxiliary equipment. From this, we conclude that significant energy 
	savings can be achieved only by temporarily disengaging redundant 
	hardware components of a base station.  Indeed, as pointed out by
  \cite{Oh2011}, reducing
the number of active base stations in periods of low traffic load
offers a huge potential for energy savings. Switching off base stations 
and introducing a cooperation between different operators in an urban scenario is
expected to lead to a reduction in energy consumption of up to 29\%
\cite{Oh2011}. This effect will become even more pronounced in the
future since, as aforementioned, the number of base stations is
expected to increase over the next years to meet the growing demand
for wireless access \cite{analysys2012}.

\subsection{Our contribution}
\label{sec:contribution}
This paper deals with the problem of minimizing the overall energy
consumption in the downlink channel of mobile (cellular) networks. 
By taking into account the energy consumed by hardware and 
auxiliary equipment, we address key shortcomings of most existing
approaches to the challenge of boosting energy
efficiency of cellular networks. The underlying
problem is of combinatorial nature because it essentially amounts to
selecting a subset of network elements corresponding to the most
energy-efficient network configuration, while
  providing the desired network coverage.  More
precisely, motivated by \cite{Zhou09}, we formulate a
combinatorial optimization problem to find a network configuration
that consumes the least amount of energy, while
satisfying traffic demands expressed
  in terms of minimum data rate requirements. In doing so, we
  balance different forms of energy consumption in an optimal manner
  by taking into account both the load-dependent energy used for
  transmission and the static energy consumed by hardware regardless
  of the actual load. Similar to \cite{Zhou09}, the technology specific
constraints are defined to capture the QoS requirements of the users.
Although our optimization framework is generic in the sense that
it can be applied to multi radio access technology
  (multi-RAT) systems by incorporating different RAT specific
  constraints, owing to the lack of space, our focus is on a single RAT 
	according to the long term evolution (LTE) standard.

The underlying combinatorial problem is in
general hard to solve so, even for networks of moderate size, we
  follow a widely-known relaxation approach (see for instance
  \cite{Liberti2004}) and relax the
problem by another problem as an
intermediate step that can be solved efficiently. More
  precisely, we apply convex relaxation
  techniques to the constraints and approximate the objective function
  by a concave function. To try to minimize the concave
  function over the convex set, we
use of a majorization-minimization (MM) technique \cite{Hunter2004}. 
Our algorithm exhibits major advantages
over other existing schemes. In particular,
  the proposed algorithm is able to find good solutions in a relatively 
	short time
so that it can even be used to utilize fluctuations in traffic
  demand on a relatively short time scale. It can further cope with a
variety of network elements such as cells, sectors or even
antennas. As a result, we can take into account different sources of energy
consumption to arrive at an energy efficient network configuration. In
  addition, the algorithm is able to balance the minimization of the
  static energy consumption, which requires the deactivation of as
  many network elements as possible, against the minimization of the
  load-dependent energy consumption, which calls for an appropriate
  load balancing to avoid highly loaded base stations. In order to
  ensure a broad range of applicability, the load-dependent energy
  consumption can be modeled as any concave or convex function of the
  load.

We also discuss how our algorithm can be extended along two directions.  First, we show how to 
incorporate techniques motivated by coordinated multi-point (CoMP) transmission \cite{1399258},
which actually makes some relaxations superfluous and leads
  to further energy savings.
Second, we use the load-based model of \cite{siomina12}
  for interference-coupled cellular networks together with the
  framework of standard interference functions \cite{yates95} to show
  that larger energy savings are possible at the cost of
  slightly increased complexity and overhead.

\subsection{Related Work}
Over recent years, some research effort has been devoted to exploiting
temporal and spatial redundancies in wireless systems for energy
savings. For instance, References
\cite{Brevis2011,Niu2010,Chiaraviglio2012} address the problem of
finding an optimal number of base stations and cell site placements so
as to minimize the overall energy consumption subject to some quality
of service (QoS) requirements of the users. Assuming a wireless
network based on time division multiple access (TDMA), the objective
of the study in \cite{Brevis2011} is to minimize the overall expected energy
consumption by optimizing the number of base stations and their
locations. The authors formulate the problem as a mixed integer
programming problem and suggest using a simplex method together with
the branch and bound algorithm. The drawback of this approach is that, due
to the TDMA assumption, the analysis does not carry over to systems
with inter-cell interference, which is one of the major challenges
faced by designers of modern wireless communication systems
\cite{Majewski2010,5876497}. Furthermore, branch and bound methods
may be slow
\cite{Joshi2009}, which excludes an application of these methods to
real-time scenarios, even if the underlying problem is of moderate
size.

References \cite{Niu2010, Zhou09} propose centralized and
decentralized algorithms for wireless communication networks to
address the problem of base station selection in the
presence of traffic load fluctuations. Although the proposed approach
seems to provide good solutions in reasonable time, it does not allow
incorporation of different sources of energy consumption,
which is of utmost importance in modern networks consisting of
hierarchical structures. In addition, the authors focus on numerical
evaluations to justify the approach. No analytical justification for
the performance of the proposed algorithms is given.

The authors of \cite{Chiaraviglio2012} argue in favor of sleep mode
techniques coupled with various network planning
schemes. A genetic algorithm is used to find
  energy-efficient network deployments, so the authors have developed a purely
	heuristic approach to put selected base stations into
a sleep mode for energy-efficient network operation. In addition to
the lack of any mathematical justification, the main shortcoming of
this work is that the proposed approach cannot incorporate
  other radio technologies other than UMTS terrestrial radio access network
  (UMTS: universal mobile telecommunications system). In contrast, as 
	mentioned before, our optimization framework is general enough to be applied to
multi-RAT scenarios, including the second, third and fourth
generations of cellular networks \cite{pollakis2013}.

 \subsection{Notation and Paper Organization}
\label{sec:preliminaries}
For a vector $\signal{x}\in \R^N$, its $i$th component is $x_i\in\R$.
Similarly, for a matrix $\ma{X}\in\R^{M\times N}$, its $(i,j)$-th
component is $x_{i,j}$. Inequalities involving vectors, such as
$\signal{x}_1\ge \signal{x}_2$, are to be understood as
component-wise inequalities. The set $\R_+$ denotes the set of non-negative real
numbers, while $\R_{++} := \R_+ \backslash \{0\}$ is the set of
positive real numbers. 

Given a matrix $\ma{X} \in \R^{M \times N}$, we use
$\tilde{\signal{x}}:=\mathrm{vec}(\ma{X})\in\R^{MN}$ to denote the
vector obtained by stacking the columns of $\ma{X}$. Note that the
entries of $\tilde{\signal{x}}$ may be confined to take values on
$[0,1]$ or $\{0,1\}$ depending on whether $\ma{X}\in[0,1]^{M\times N}$
or $\ma{X}\in\{0,1\}^{M\times N}$.
\begin{definition}[$l_0$-norm]
\label{def:l0_metric}
For any vector $\signal{x} \in \R^N$ and matrix $\ma{X}\in\R^{M\times
  N}$, \added{their $l_0$-norms} $|\signal{x}|_0$ and $|\signal{X}|_0$
\changed{are equal to}{is used to denote} the number of nonzero
elements of $\signal{x}$ and $\ma{X}$, respectively. For a scalar $x
\in \R$, $| x |_0:=1$ if $x\ne 0$ and $| x |_0:=0$
otherwise.\footnote{\added{Although the $l_0$-norm is not a norm, we
    use the term ``norm'' as it is a common practice in \removed{the compressed
    sensing} literature.}} 
\end{definition}

The remainder of this paper is organized as
follows. Section~\ref{sec:model} introduces the underlying system
model and in section~\ref{sec:model_problem} we outline the general 
problem to solve. 
In section~\ref{sec:SparseOptiEnergyEffi} the proposed algorithm to find
solutions for our optimization problem is derived based on a 
worst-case inter-cell interference assumption.
Section~\ref{sec:sif} presents how to explicitly take into account a more
realistic inter-cell interference model. We present empirical
evaluations of the proposed algorithm in section~\ref{sec:numericalResults}.

\section{System Model}
\label{sec:model}
We consider the downlink channel of a multi-cell LTE network with an
established network topology and a central network controller. The central
network controller is responsible for collecting measurements,
executing the proposed algorithm, and propagating updated network
configuration parameters throughout the network. 
We assume that the network consists of $L$ base stations. Each base station
has multiple sectors (called cells in the following), and we denote
the set of cells
belonging to base station $l$ by $\sets_l$. The set of all base stations
is denoted by $\setl$, and we use
$\setm:= \cup_{l\in \setl} \sets_l$ to denote the set of all $M$ cells in the
network. The cell deployment is assumed to be dense enough
so that coverage areas of different cells overlap. This
implies that users can be served by different neighboring cells.
\subsection{Ensuring coverage via test points}
\label{sec:model_coverage}
In order to ensure the desired coverage anytime and everywhere in the
considered area, we impose coverage constraints by adopting the
concept of test points \cite{Capone03}. 
\begin{definition}[Test point]
\label{def:test_point}
A test point (TP) is a centroid of a pre-defined subarea that represents an
aggregated QoS requirement resulting from individual QoS demands of
all potential users in this subarea.\footnote{\added{A test point
    becomes a user if it represents a QoS requirement of one
    particular user, in which case the subarea is a point corresponding
    to the position of this user.} } Without loss of generality, we
assume $N$ TPs with the set of all TPs denoted by
$\setn:=\left\{ 1,2,...,N\right\}$.
\end{definition} 

A consequence of this definition is that small-scale fluctuations in
QoS demand at the user level are averaged out at the TPs. 
These small-scale fluctuations must be compensated by the
lower layers of the protocol stack (e.g. through adaptive modulation
or coding).
\begin{assumption}
 \label{as:testpoint_rate}
 The QoS requirement for a TP corresponds to the aggregated
 expected traffic over the respective area per unit
 time. This traffic requirement is expressed in terms of the minimum
 required data rate per TP. 
\end{assumption}
\begin{assumption}
\label{as:user_req}
If the minimum rate requirement of TP $j$ is met, so are the
requirements of the users in the associated subarea.\footnote{The
  smaller the area represented by each TP, the better is this
  approximation. However, smaller areas imply an increased number
  of TPs, and the computational complexity of the proposed
  algorithm grows.}
\end{assumption}
For services with no explicit data rate requirements (e.g. voice
calls), we assume that they can be supported if a minimum data rate
per service request is ensured. By Assumption \ref{as:testpoint_rate},
each TP $j\in\setn$ is assigned rate requirement $r_j$, and we
collect the rate requirements of all TPs in the vector
$\signal{r} = \left[ r_1,r_2, \ldots, r_N \right] \in \R_{++}^N$.
In general, a TP can be assigned to any cell, and
an assignment should be understood as follows. If TP
$j\in\setn$ is assigned to cell $i\in\setm$, then all users in
the respective subarea associated with TP $j$ are served by
cell $j$. The assignment of the TPs to the cells 
is subject to optimization in this paper.  We use
$\ma{X}=[x_{i,j}] \in \{0,1\}^{M \times N}$ to denote the assignment
matrix where $x_{i,j}=1$ if TP $j$ is assigned to cell
$i$ and $x_{i,j}=0$ otherwise.
\begin{assumption}
  \label{as:test_one_base}
While each TP is assigned to exactly one cell, each
cell can serve multiple TPs, and the set of TPs
served by cell $i$ under assignment $\ma{X}$ is denoted by $\setn_{i}(\ma{X}) \subset \setn$.\par
\end{assumption}
We point out that this assumption has been widely used in previous studies
\cite{Capone03,Zhou09,Niu2010}, and it is valid throughout the paper
except for Section \ref{sec:comp}, where it is shown how to include
scenarios in which each TP can be served by multiple cells.
Note that if $\setn_i(\ma{X})=\emptyset$ for some $i\in\setm$, then cell
$i$ can be deactivated for energy savings because no TP
is assigned to cell $i$.  In contrast, if $\setn_i(\ma{X}) \neq
\emptyset$, then cell $i$ is active, and each TP
connected to it induces some amount of \textit{cell load}.
\begin{definition}[Cell Load]
\label{def:cell_load}
Given the assignment $\tilde{\signal{x}}:=\mathrm{vec}(\ma{X})$, the load of cell $i$,
denoted by $\rho_i(\tilde{\signal{x}})\in[0,1]$ or simply $\rho_i$ for notational simplicity, 
is defined to be the ratio of the number of resource blocks requested by TPs
served by cell $i\in\setm$ to the total number of resource
blocks $B_i$ available at this cell.\footnote{Note that $B_i$
  can also be interpreted as the total bandwidth available at cell
  $i$, in which case $\rho_i$ is expressed in terms of the
  fraction of required and available bandwidth.}
\end{definition}
We use $\signal{\rho}:=[\rho_1,\ldots,\rho_M]^T \in [0,1]^M$ to denote
the vector of all cell loads. From the definition of cell load, we have the following:
\begin{fact}
\label{fac:load_positive}
The load at cell $i$ satisfies $\rho_i>0$ if and only if (iff) cell $i$ serves at least one
TP.
\end{fact}

\subsection{Spectral efficiency and resource usage}
\label{sec:model_load}
The optimal assignment of TPs to cells is strongly
influenced by the spectral efficiency of the corresponding links. For
the analysis in this paper, we adopt an OFDMA-based model for the
spectral efficiency that is widely used in the literature
\cite{Majewski2010, Morgensen07,6576482}. The spectral efficiency also depends on radio propagation
properties. Therefore, we associate to each TP a path-loss
vector and write the path-loss vectors of all TPs as columns
of the path-loss matrix $\ma{G}=[g_{i,j}]\in\R_{++}^{M \times N}$,
where $g_{i,j}$ captures the long-term path loss and shadowing effects
for a radio link from cell $i$ to TP $j$.
\begin{assumption}[Reliable path-loss estimates]
 \label{as:CSI_available}
 A reliable estimate of $\ma{G}$ is available at the central network
 controller.
\end{assumption}

\begin{remark}
\label{rem:robust estimate}
The problem of reliable estimation and tracking of the path-loss
matrix is out of the scope of the paper. However, the matrix captures
only long-term fading effects, so reliable estimates of $\ma{G}$ can
be obtained and tracked in practice. Promising algorithmic solutions
to this estimation problem are for instance presented in
\cite{KCVSY14aoro}. Moreover, in network planning problems, knowledge
of $\signal{G}$ is a very common assumption in the literature
\cite{siomina12,Majewski2010,Capone03}.
\end{remark}

Now we are in a position to define the
signal-to-interference-noise-ratio (SINR) $\gamma_{i,j}:\R_+^M \to
\R_+$ between cell $i\in\setm$ and TP $j\in\setn$ by
\cite{siomina12,Majewski2010}:
\begin{equation}
\label{eq:sinr_load}
\gamma_{i,j}(\signal{\rho}) = \frac{P_i ~ g_{i,j}}{\sum_{k\in\setm\backslash\{i\}}{P_k ~ g_{k,j} ~ \rho_k}+\sigma^2},
\end{equation}
where $P_i>0$ is the transmit power per resource block of cell
$i$ and $\sigma^2>0$ is the noise power per resource block.
Accordingly, the link spectral efficiency $\omega_{i,j}: \R_+^M
\to \R_+$ (in bits per resource block\footnote{A resource block is
  defined as a portion of the available time-frequency plane spanning
  a number of consecutive OFDM symbols in the time domain over a
  number of sub-carriers in the frequency domain.})  for the link
from cell $i$ to TP $j$ is given by \cite{Morgensen07}
\begin{equation}
\label{eq:specEffPL}
\omega_{i,j}(\signal{\rho}) = \eta^{\text{BW}}_{i,j} \log_2 \Bigl(1+ 
\frac{\gamma_{i,j}(\signal{\rho})}{\eta^{\text{SINR}}_{i,j}} \Bigr) 
\end{equation}
where $\eta_{i,j}^\mathrm{BW}\in\R_{++}$ and
$\eta_{i,j}^\mathrm{SINR}\in\R_{++}$ are suitably chosen constants,
referred to as bandwidth and SINR efficiency, respectively. These
constants depend on the overall system design, which includes the
choice of scheduling protocols and multi-antenna techniques. The
choice of these constants has no impact on our results, so they are
assumed to be arbitrary and fixed throughout the paper. For realistic
values of these constants, we refer the interested reader to
\cite{Majewski2010,Morgensen07}.

From (\ref{eq:specEffPL}),
we can easily see that the necessary number of resource blocks
$b_{i,j}$ at cell $i$ to serve TP $j$ with data rate
$r_j$ is equal to $b_{i,j} =
\frac{r_j}{\omega_{i,j}(\signal{\rho}) }>0$. In addition, following
Definition \ref{def:cell_load}, the load at cells can be
computed by the following system of non-linear equations
\begin{equation}
\rho_i = \sum_{j \in \setn_i(\ma{X})} \frac{b_{i,j}}{B_i} = \sum_{j \in \setn_i(\ma{X})} \frac{r_{j}}{B_i \omega_{i,j}(\signal{\rho})},\, i \in \setm.
\label{eq.lteqos}
\end{equation}
\begin{remark}
\label{rem:signaling}
In practice, cells need to reserve some fraction of their
resource blocks for signaling.  If cell $i$ has $B^*_i$
resource blocks in total, and it needs to reserve $a_i>0$ of its resource
blocks for signaling, then the resource blocks at cell $i$
available for allocation to TPs are $B_i = B^*_i - a_i$.
\end{remark}

For a fixed assignment $\ma{X}$ cell load
$\signal{\rho}$ in \refeq{eq.lteqos} can be efficiently computed by means of fixed-point algorithms (c.f. Sec.~\ref{sec:sif}).
However, the assignment of TPs to cells
is the main subject of our optimization problem and thus we cannot evaluate \refeq{eq.lteqos} easily. In order to keep the complexity of the optimization problem tractable, we lower bound the spectral efficiency. 
\begin{assumption}[Worst-Case Interference]
\label{as:worst_case_inter}
We have the worst-case interference scenario if all cells are
fully loaded, i.e. $\signal{\rho} = \signal{1}$.
\end{assumption}

Unless otherwise stated, we assume the worst-case interference, which
results in a lower bound on the link spectral efficiency
$\omega_{i,j}(\signal{\rho})\geq\tilde{\omega}_{i,j}:=\omega_{i,j}(\signal{1})$ for every $\signal{\rho}\in [0,1]^M$. In general, 
this bound diminishes gains in energy savings when taking into account the energy consumption of hardware, and we show in
Section \ref{sec:sif} how to incorporate the actual link spectral
efficiency to improve the energy savings. Nevertheless, having fully loaded cells as in Assumption~\ref{as:worst_case_inter} is desirable because it has been
proven in \cite{HoYuan15} that full load (i.e. $\signal{\rho} = \signal{1}$) is optimal with respect to the transmit energy consumption (see also \cite{cavalcante2014}).
\begin{remark}
\label{rem:worst_interference}
The worst-case interference assumption cannot exploit the full
potential for energy savings, but the assumption is of high practical
relevance because it is an effective way to avoid coverage holes as a
result of deactivating cells based, for instance, on imperfect
information.
\end{remark}
\subsection{Energy consumption model}
\label{sec:model_energy}
In contrast to most work in literature, we consider a model for the energy
consumption of a base station and its cells that
takes into account not only the cell load-dependent transmit energy
radiated by antennas, but also the remaining sources of energy
consumption that are independent of the cell load as long as the cell/base
station is \emph{active}.
\begin{definition}[Active base station/cell]
\label{def:active_BS}
\label{def:active_cell}
Consider a particular base station $l\in\setl$ and its cells $i\in\sets_l$. 
Let $\rho_i\in[0,1]$ be the load of cell $i$. We say that
a cell $i$ is \emph{active} iff $\rho_i>0$ and that base station $l$ is 
\emph{active} iff one of its cells is \emph{active}, 
i.e. $\sum_{i\in\sets_l}\rho_i>0$. If a cell or base station is not active 
it is said to be \emph{inactive}.
\end{definition}
With Definition~\ref{def:active_cell} we are in the position to define the energy consumption of a 
base station.
\begin{definition}[Energy consumption]
\label{def:energy_cons}
Given a TP assignment $\ma{X}$ inducing a cell load $\signal{\rho}$, the energy consumption
$E_l(\signal{\rho})\geq 0$ of base station $l$ is defined to be the
power that the respective base station consumes per unit
of time, where $E_l(\signal{\rho})=0$ iff base station $l$ is inactive. 
\end{definition}
The function $E_l(\signal{\rho})$ depends on the hardware setup of the base station, but it can be split into three parts in general:
\begin{enumerate}[(i)]
\item The static energy consumption of the base station $c_l>0$ 
(due to shared hardware between sectors e.g. cooling, power supply, etc.), 
\item The static energy consumption $e_i>0$ ($i\in\sets_l$) of its \emph{active} cells
	(e.g. due to power amplifiers, signal processing units, etc.), 
	and
\item the load-dependent dynamic energy consumption of its \emph{active} cells
  $f_{i}(\rho_i)$ ($i\in\sets_l$), where $f_i:[0,1]\to\mathbb{R}_+$ is a
  given continuous function relating the energy consumption to the
  corresponding cell load.
\end{enumerate}

By these definitions and Fact \ref{fac:load_positive}, $E_l(\signal{\rho})$ is a
discontinuous function of the cell load, and we have
\begin{equation*}
E_l(\signal{\rho}) =\begin{cases} 
0 & \text{cells $i\in\sets_l$ serve no TP}\\
c_l + \sum_{i\in\sets_{l, \text{active}}} e_i + f_{i}(\rho_i) & \text{otherwise}\,,
\end{cases}
\end{equation*}
where $\sets_{l, \text{active}}\subset \sets_{l}$ is the set of of active cells of base station $l$.
Therefore, the total energy consumption in a network, which is the
accumulated energy consumption of all active base stations, yields
\begin{equation}
\label{eq:totalNetEnergy}
E(\signal{\rho}) = \sum_{l\in\setl} E_l(\signal{\rho})
= \sum_{l \in \setl}{  \left(c_l\left|\mathsmaller{\sum_{i\in\sets_l}} \rho_i \right|_0 + \sum_{i\in\sets_l}
\left( e_i | \rho_i |_0 + f_{i}(\rho_i)\right)\right)}.
\end{equation}
For concreteness, we make the following assumption throughout the
paper (see also Remark \ref{rem:dynamicload_convex})
\begin{assumption}[Concave dynamic energy consumption]
\label{as:dynamic_energy}
$f_{i}:[0,1]\to\RN \, (i\in\setm)$, is concave and continuously
differentiable.
\end{assumption}
In particular, this assumption is satisfied by a linear dependency of the base station energy consumption and the cell load reported in current studies such as \cite{earthd23}.
\begin{remark}
\label{rem:dynamicload_convex}
In fact, the load-dependent dynamic energy consumption can also be
assumed to be a convex function of the load. Moreover, we could even
assume that it is a sum of convex and concave functions. The
optimization framework presented in this paper can be
straightforwardly extended to cover these cases.
\end{remark}

\section{Problem Statement}
\label{sec:model_problem}
Spatio-temporal redundancies in coverage and capacity resulting from
day-time fluctuations in traffic demand present great opportunities
for energy savings by deactivating redundant cells at times of
relatively low traffic demand.
Indeed, if the traffic demand decreases, some or all entries of the
rate requirement vector $\signal{r}\in\mathbb{R}^N_{++}$ become
relatively small, which can be utilized to reduce the total energy
consumption by minimizing the cost function in
(\ref{eq:totalNetEnergy}) subject to different constraints that follow
from the system model and (\ref{eq.lteqos}). Formally, the problem
under consideration can be stated as follows (note that the complete
set of equations is referred to as (\ref{eq:problem_original})):
\begin{subequations}
\label{eq:problem_original}
\begin{align}
\mbox{min. } &\lefteqn{\sum_{l \in \setl} \left(c_l\left|\sum_{i\in\sets_l} \rho_i \right|_0 + \sum_{i\in\sets_l} \left(e_i| \rho_i |_0 + f_{i}(\rho_i)\right)\right)} \label{eq:p1.0}\\
  \mbox{s. t.: } & \sum_{j \in \setn} \frac{r_j}{B_i ~ \tilde{\omega}_{i,j}} ~ x_{i,j} = \rho_i & i \in \setm \label{eq:p1.4}\\
& \sum_{i \in \setm} x_{i,j} = 1 &  j  \in \setn \label{eq:p1.5}\\
& \rho_i \in \left[0,1\right] & i\in \setm \label{eq:p1.56}\\
& x_{i,j} \in \left\{ 0,1 \right\} & i \in \setm, j \in \setn  \label{eq:p1.6}\,,
\end{align}
\end{subequations}
where the optimization variables are $x_{i,j}$ and $\rho_i$ ($i \in
\setm, j \in \setn$).  In particular, Assumption
\ref{as:test_one_base} is captured by (\ref{eq:p1.5}) together with
(\ref{eq:p1.6}).
Constraints \refeq{eq:p1.4} and (\ref{eq:p1.56}), in contrast, ensure
that the cell load is in accordance with Definition
\ref{def:cell_load}. \par
To ensure feasibility of the above problem and to show the
effectiveness of our approach, we consider scenarios where the rate
requirements of TPs are sufficiently low for a reasonable
amount of redundancies that allow for deactivation of cells.
Moreover, if
the traffic requirements in the system are sufficiently low or the
number of cells is sufficiently large, $\signal{\rho}^\star$
is expected to be \emph{sparse} with zero entries specifying cells
that can be deactivated.

\section{Energy-Efficiency Optimization}
\label{sec:SparseOptiEnergyEffi}
The difficulty of problem (\ref{eq:problem_original}) lies in
its combinatorial nature.
In fact, it can be shown that the problem is related to the classical
bin-packing problem,
which is known to be NP-hard
\cite{cavalcante2013energy}. Consequently, the complexity is expected
to grow exponentially with the number of cells. On the
positive side, problem (\ref{eq:problem_original}) has a
special structure that can be exploited by majorization-minimization
techniques \cite{Hunter2004}, which have been widely used in recent
years to tackle various problems in compressed sensing
\cite{Candes2008} and machine learning \cite{Sriperumbudur2011}.

Instead of finding a global solution to
(\ref{eq:problem_original}), we will pursue a less ambitious goal. We
apply the majorization-minimization techniques mentioned above to
develop a low-complexity anytime algorithm that has a strong
analytical justification. This algorithm is expected to provide good
results (in terms of low energy consumption) with low-complexity. To this end, we reformulate problem
(\ref{eq:problem_original}) to pose it in a more tractable
form. First, we observe that each load $\rho_i$ is, in fact, a function of
$\ma{X}$ (c.f. Definition \ref{def:cell_load} and \refeq{eq:p1.4}). We
can therefore modify the problem to have only $\ma{X}$ as an
optimization variable.  Recall that, if at least one TP
is served by cell $i$ ( i.e., $\sum_{j\in \setn} x_{i,j}\geq
1$), then it follows from Fact \ref{fac:load_positive} that the cell
load at cell $i$ is non-zero and $| \rho_i |_0 = 1$. Hence,
the objective function in (\ref{eq:p1.0}) can be equivalently written
as
\begin{equation}
\label{eq.1.0.1}
\begin{split}
&\sum_{l \in \setl}{  \left(c_l\left|\sum_{i\in\sets_l} \rho_i \right|_0 + \sum_{i\in\sets_l} \left(e_i \left| \rho_i \right|_0 + f_{i}(\rho_i)\right)\right)}\\
&=\sum_{l \in \setl}{  \left(c_l\left|\sum_{i\in\sets_l}\sum_{j\in \setn} x_{i,j} \right|_0 + 
\sum_{i\in\sets_l} \left(e_i \left| \sum_{j\in \setn} x_{i,j} \right|_0 + f_{i}(\rho_i)\right)\right)}\\
&=\sum_{l \in \setl}{  \left(c_l\left|\signal{t}_l^T \tilde{\signal{x}}  
\right|_0 + \sum_{i\in\sets_l} \left(e_i \left| \signal{s}_i^T \tilde{\signal{x}}  \right|_0 + f_{i}(\rho_i)\right)\right)}
\end{split}
\end{equation}
where $\signal{s}_i := \text{vec}(\ma{S}_i)$ with $\ma{S}_i\in \{0,1
\}^{M \times N}$ being a matrix of zeros, except for its $i$th row,
which is a row of ones. Similarly, $\signal{t}_l := \text{vec}(\ma{T}_l)$ with $\ma{T}_l\in \{0,1
\}^{M \times N}$ is a matrix of zeros, except for its rows $i\in\sets_l$,
which are rows of ones.

\begin{definition}
\label{def:tansform_dyn_load}
Given the assignment $\tilde{\signal{x}}$ and the load dependent energy consumption $f_i(\rho_i(\tilde{\signal{x}}))$ of cell $i$ with $\rho_i(\tilde{\signal{x}}) =\sum_{j \in \setn} \frac{r_j}{B_i \tilde{\omega}_{i,j}}\,x_{i,j} $ (c.f. \refeq{eq:p1.4}), we define the function
$\tilde{f}_i:[0,1]^{NM} \to \R_+ : \tilde{\signal{x}} \mapsto f_i(\sum_{j \in \setn} \frac{r_j}{B_i \tilde{\omega}_{i,j}}\,x_{i,j} )$.
\end{definition}

Considering Definition \ref{def:tansform_dyn_load} and using $\rho_i
\leq 1$ (see Definition \ref{def:cell_load}) in \refeq{eq:p1.4}, we
arrive at an equivalent problem given by
\begin{subequations}
\label{eq:problem_equiv}
 \begin{align}
\mbox{min. } &\lefteqn{\sum_{l \in \setl}{  \left(c_l\left|\signal{t}_l^T \tilde{\signal{x}}  \right|_0 + \sum_{i\in\sets_l}\left( e_i \left| \signal{s}_i^T \tilde{\signal{x}} \right|_0 + \tilde{f}_{i}(\tilde{\signal{x}} )\right)\right)}} \label{eq:p2.0}\\
  \mbox{s. t.: } & \sum_{j \in \setn} \frac{r_j}{B_i ~ \tilde{\omega}_{i,j}} ~ x_{i,j} \leq 1 & i \in \setm \label{eq:p2.1}\\
& \sum_{i \in \setm} x_{i,j} = 1 &  j  \in \setn \label{eq:p2.2}\\
& x_{i,j} \in \left\{ 0,1 \right\} & i \in \setm, j \in \setn  \label{eq:p2.3},
\end{align}
\end{subequations}
where the assignment variables $x_{i,j}$ ($i \in \setm, j \in \setn$) are the only optimization
variables.

\subsection{Problem relaxation}
\label{sec:SparseOptiEnergyEffi_relaxation}
To obtain an optimization problem that is computationally tractable, we first relax the binary
constraint \refeq{eq:p2.3} to\footnote{This relaxation together
  with \refeq{eq:p2.2} leads to a communication scenario where
  multiple cells serve one TP. A more detailed
  discussion on the implications is presented in Sec.~\ref{sec:comp}}
 \begin{equation}
 \label{eq:relaxedCo}
  x_{i,j} \in \left[ 0,1 \right], \forall i\in\setm, \forall j\in\setn.
\end{equation}
The above makes all constraints convex, so now the only problem is the
objective function, which is not continuous due to the
$l_0$-norm. We also note that by Assumption \ref{as:dynamic_energy} and Definition \ref{def:tansform_dyn_load}, the load-dependent term
$\tilde{f}_i(\tilde{\signal{x}})$ in the objective function \refeq{eq:p2.0} is concave
and continuously differentiable for $\tilde{\signal{x}}\in[0,1]^{NM}$
since these properties are preserved under a composition with a linear
function \cite{baus11, boyd}.
To address the non-continuity of the $l_0$-norm, we consider
the following relation \cite{Candes2008}:
\begin{empheq}{align}
\forall_{\signal{z}\in \R^K}\; |\signal{z}|_0=\lim_{\epsilon\to 0}\sum_{k=1}^K \dfrac{\log(1+|z_k| ~ \epsilon^{-1})}{\log(1+\epsilon^{-1})}.
\label{eq:l0reform}
\end{empheq}
By using \refeq{eq:l0reform} and the non-negativity of $\signal{s}_i$, $\signal{t}_i$, $\tilde{\signal{x}}$, the cost function in \refeq{eq:p2.0} can be equivalently written as
\begin{equation}
\begin{split}
&\sum_{l \in \setl}{  \left(c_l\left|\signal{t}_l^T \tilde{\signal{x}}  \right|_0 + \sum_{i\in\sets_l} e_i \left| \signal{s}_i^T \tilde{\signal{x}} \right|_0 + \tilde{f}_{i}(\tilde{\signal{x}} )\right)} \\
&=\lim_{\epsilon\to 0}\sum_{l \in \setl} \left(c_l\dfrac{ \log(1+\epsilon^{-1} ~ \signal{t}_l^T \tilde{\signal{x}})}{\log(1+\epsilon^{-1})} 
+ \sum_{i\in\sets_l} \left(e_i \dfrac{ \log(1+\epsilon^{-1} ~ \signal{s}_i^T \tilde{\signal{x}})}{\log(1+\epsilon^{-1})} + \tilde{f}_{i}(\tilde{\signal{x}} )\right)\right).
\end{split}
\label{eq:objlog}
\end{equation}
We can therefore obtain an approximation to problem
(\ref{eq:problem_original}) by replacing the objective function by the
right-hand side of \refeq{eq:objlog} for a sufficiently small but
fixed $\epsilon>0$. More precisely, for some $\epsilon>0$, the
objective is to find a matrix $\ma{X}\in[0,1]^{M\times N}$
or, equivalently, a vector
$\tilde{\signal{x}}=\mathrm{vec}(\ma{X})\in[0,1]^{NM}$ that solves the
following problem
\begin{subequations}
\label{eq.smm_prob}
\begin{empheq}{align} 
\mbox{min. } &\sum_{l \in \setl}  \left(c_l\dfrac{ \log(1+\epsilon^{-1} ~ \signal{t}_l^T 
\tilde{\signal{x}})}{\log(1+\epsilon^{-1})}
+ \sum_{i\in\sets_l} \left( e_i \dfrac{ \log(1+\epsilon^{-1} ~ \signal{s}_i^T \tilde{\signal{x}})}{\log(1+\epsilon^{-1})} 
+ \tilde{f}_{i}(\tilde{\signal{x}} )\right)\right) \label{eq.1.1.0}\\
  \mbox{s. t.: } &  \sum_{j \in \setn} \frac{r_j}{B_i ~ \tilde{\omega}_{i,j}} ~ x_{i,j} \leq 1& i\in \setm \label{eq.1.1.1}\\
  & \sum_{i \in \setm} x_{i,j} = 1, &   j \in \setn \label{eq.1.1.2}\\
   & x_{i,j} \in \left[ 0,1 \right] & i \in \setm, j \in \setn.\label{eq.1.1.3}
\end{empheq}
\end{subequations}
Solving problem \refeq{eq.smm_prob} is not straightforward because we need to
\textit{minimize} a non-convex function over a convex
set. Fortunately, Reference \cite{Candes2008} presents an optimization
framework based on the majorization-minimization (MM) algorithm
\cite{Hunter2004} to handle problems of this type. The framework can
be used to decrease the value of the objective function in a
computationally efficient way. For completeness, we the reader can find some details of the MM algorithms in the appendix.

\subsection{Majorization-minimization (MM) algorithm}
\label{sec:mm_algorithm}
For notational convenience, we define $\hat{c}_l :=
\frac{c_l}{\log(1+\epsilon^{-1})}$ and $\hat{e}_i :=
\frac{e_i}{\log(1+\epsilon^{-1})}$, and we use these definitions in \refeq{eq.1.1.0} to
simplify the objective function (ignoring unnecessary constants):
\begin{equation}
\label{eq:objective}
h:\mathcal{X}\to\mathbb{R}:\; \tilde{\signal{x}} \mapsto
h(\tilde{\signal{x}})=\sum_{l\in \setl} \left( \hat{c}_l \log(\epsilon+ \signal{t}_l^T \tilde{\signal{x}}) \right) 
+ \sum_{i\in\setm} \left(\hat{e}_i \log(\epsilon+ \signal{s}_i^T \tilde{\signal{x}}) + \tilde{f}_{i}(\tilde{\signal{x}} ) \right),
\end{equation}
where $\mathcal{X}\subset \R^{M N}$ is the closed convex set of points
satisfying the constraints \refeq{eq.1.1.1}-\refeq{eq.1.1.3} and we have used the fact that $\setm = \cup_{l\in\setl}\sets_l$. Since
$\tilde{f}_{i}$ is concave and continuously differentiable by
Assumption \ref{as:dynamic_energy}, so is the function in
(\ref{eq:objective}) for any $\epsilon>0$. Therefore, according to the
explanations in the appendix, we can use the following function
\begin{equation*}
g:\mathcal{X}\times\mathcal{X}\to\mathbb{R}: \, (\signal{x},\signal{y})\mapsto h(\signal{y})+\nabla h(\signal{y})^T(\signal{x}-\signal{y})
\end{equation*}
as a majorizing function of \refeq{eq:objective}, where the gradient 
can be easily calculated:
\begin{equation}
 \nabla h(\tilde{\signal{x}})=\sum_{l\in \setl} \hat{c}_l \dfrac{1}{\epsilon+\signal{t}_l^T \tilde{\signal{x}}} + \sum_{i\in\setm}\Bigl(\hat{e}_i \dfrac{1}{\epsilon+\signal{s}_i^T \tilde{\signal{x}}} + \nabla \tilde{f}_{i}(\tilde{\signal{x}} ) \Bigr).
\label{eq:gradient}
\end{equation}
Thus, updates of the MM algorithm take the form (see the appendix)
\begin{equation}
\label{eq.mmit}
\begin{split}
 &{\tilde{\signal{x}}}^{(n+1)} \in \arg\min_{\tilde{\signal{x}}\in\mathcal{X}}g({\tilde{\signal{x}}},{\tilde{\signal{x}}}^{(n)}) \\ 
&= \arg\min_{\tilde{\signal{x}}\in\mathcal{X}} \sum_{l\in \setl} \hat{c}_l \dfrac{\signal{t}_l^T \tilde{\signal{x}}} {\epsilon+\signal{t}_l^T \tilde{\signal{x}}^{(n)}}
+ \sum_{i\in\setm}\left(\hat{e}_i \dfrac{\signal{s}_i^T \tilde{\signal{x}}} {\epsilon+\signal{s}_i^T \tilde{\signal{x}}^{(n)}} + \nabla \tilde{f}_{i}(\tilde{\signal{x}}^{(n)} )^T \tilde{\signal{x}}\right)
\end{split}
\end{equation}
for some feasible starting point\footnote{In our experience a good
  starting point is derived from a feasible assignment matrix obtained
  by connecting each TP to the cell providing the
  strongest received signal strength.}
$\tilde{\signal{x}}^{(0)}\in\mathcal{X}$. In words, the MM algorithm
solves iteratively a sequence of convex optimization problems. For the chosen majorizing function, the problem to be solved in every iteration is a
linear programming problem (LP), which can be typically solved efficiently with standard optimization tools.

As discussed in the appendix, the sequence
$\{\tilde{\signal{x}}^{(n)}\}_{n\in\Natural}\subset\mathcal{X}$ for
some $\tilde{\signal{x}}^{(0)}\in\mathcal{X}$ generated by
(\ref{eq.mmit}) produces a non-increasing sequence
$\{h(\tilde{\signal{x}}^{(n)})\}_{n\in\Natural}$ of objective values.
Therefore, as $n \to \infty$, we expect the corresponding sequence of
assignment matrices $\{\signal{X}^{(n)}\}_{n\in\Natural}$ (note that
$\tilde{\signal{x}}^{(n)} =: \text{vec}(\ma{X}^{(n)}) $) to evolve
towards network configurations with low energy consumption.

We stop the algorithm if the improvements in the objective value are
small enough in the sense that for some sufficiently small
$\epsilon^\star>0$, the following condition is met
\begin{empheq}{align}
\label{eq:convergenceK}
h \bigl(\tilde{\signal{x}}^{(n)}\bigr) - h \bigl(
\tilde{\signal{x}}^{(n+1)} \bigr) \leq \epsilon^\star\,.
\end{empheq}

Upon termination, the resulting assignment matrix $\ma{X}^{(n)} \in
\left[0,1\right]^{M \times N}$ needs to be mapped to a matrix
$\ma{X}^\star \in \left\{0,1\right\}^{M \times N}$ in order to obtain
a feasible point to the problem in (\ref{eq:problem_original}). For
this purpose, we use the heuristic described in
Alg.~\ref{alg:userAllocation}. The main idea is as follows. We start
by rounding the entries $x^{(n)}_{i,j}$ to the closest integer, and
then we check if the obtained assignment matrix is part of the set
$\mathcal{X}$. \changed{Otherwise}{If this heuristic fails}, we
activate additional cells and connect TPs to them. By
using the standard LP solver of CPLEX, in our simulations most entries
of the matrix $\ma{X}^{(n)} \in \left[0,1\right]^{M \times N}$ are
typically either zero or one, so the rounding operation rarely results
in a violation of a constraint (but we emphasize that this is not
guaranteed to be true in general).  \par
\begin{algorithm}
  \caption{Heuristic to map $\left[0,1\right]^{M \times N} \rightarrow \left\{0,1\right\}^{M \times N}$}
\label{alg:userAllocation}
\begin{algorithmic}[1]
  \REQUIRE $\ma{X}^{(n)}$, $\setn$, $\setm$, set of constraints $\mathcal{X}_2$ representing  \refeq{eq.1.1.1} and \refeq{eq:p1.6}
  \ENSURE final assignment matrix $\ma{X}^\star$
\STATE initialize: set of assigned TPs $\set{A} = \emptyset$ and final assignment matrix $\ma{X}^\star = \ma{0}$.
\FORALL{ $i\in\setm, j\in\setn$}
\IF{$x^{(n)}_{i,j}\in\{1\}$}
\STATE $ x_{i,j}^\star =x^{(n)}_{i,j}$  and $\set{A} = \set{A} \cup \left\{ j\right\}$.
\ENDIF
\ENDFOR
\STATE Define set $\set{B} = \left\{x^{(n)}_{i,j} \in \left(0,1\right) \left| \forall i \in \setm, \forall j \in \setn \backslash \set{A} \right.\right\}$.
\WHILE{ $\set{B} \neq \emptyset$ }
\STATE $(i,j) = \argmax_{i,j}\{ \set{B} \}$
\IF{$x_{i,j}^\star :=1 \rightarrow  \ma{X}^\star \in \mathcal{X}_2$ }
\STATE $x_{i,j}^\star =1$ and  $\set{A} = \set{A} \cup \left\{ j \right\}$.
\STATE $\set{B} = \set{B} \backslash \{x^{(n)}_{i,j} \left| \forall i \in \setm \right. \}$
\ELSE
\STATE $\set{B} = \set{B} \backslash \{x^{(n)}_{i,j}\}$
\ENDIF
\ENDWHILE
\FORALL{ $j \notin \set{A}$}
\STATE activate closest non-active cell $i$ which yields $x_{i,j}^\star :=1 \rightarrow  \ma{X}^\star \in \mathcal{X}_2$ and assign $x_{i,j}^\star =1$.
\STATE $\set{A} = \set{A} \cup \left\{ j \right\}$.
\ENDFOR
\end{algorithmic}
\end{algorithm}
For convenience, we summarize the complete approach in
Alg.~\ref{alg:networkReconfiguration}.
\begin{algorithm}
  \caption{Network reconfiguration for improved energy efficient operation}
\label{alg:networkReconfiguration}
\begin{algorithmic}[1]
  \REQUIRE set of TPs, set of cells, constraints 
  \ENSURE optimized network configuration according to $\ma{X}^{\star}$.
\STATE initialize ${\ma{X}}^{(0)}$ with a feasible point.
\REPEAT
\STATE compute ${\tilde{\signal{x}}}^{(n)}$ by solving 
(\ref{eq.mmit})
\STATE increment $n$
\UNTIL (\ref{eq:convergenceK}) is valid
\STATE use Alg.~\ref{alg:userAllocation} to map $\ma{X}^{(n)}$
to $\ma{X}^{\star} \in \left\{0,1\right\}^{M \times N}$
\STATE connect the TPs to cells according to $\ma{X}^{\star}$.
\STATE deactivate all cells no TP is connected to.
\end{algorithmic}
\end{algorithm}

\subsection{Serving a test point with multiple cells}
\label{sec:comp}
By Assumption \ref{as:test_one_base}, 
each TP is restricted to be served by exactly one cell.
This strict limitation introduces the non-convex constraint
(\ref{eq:p1.6}) to the optimization problem in
(\ref{eq:problem_original}), which motivates the relaxation
(\ref{eq:relaxedCo}) and the heuristic mapping introduced in
Alg.~\ref{alg:userAllocation}. To avoid these heuristic approaches for
which we are not guaranteed to find solutions, we assume in this
section that each TP can be served by multiple cells.
This assumption is implemented by using \refeq{eq:relaxedCo} directly
instead of \refeq{eq:p1.6}. As a result, there is no need for any
relaxations of the constraints or the use of heuristic mappings such
as that in Alg.~\ref{alg:userAllocation}. We only need to approximate
the cost function as done in \refeq{eq.1.1.0} and apply the MM
algorithm to the resulting optimization problem, and we note that
these operations have a strong analytical justification.

The assumption of multiple cells serving \changed{one}{each}
TP has a practical interpretation when considering Definition
\ref{def:test_point}. It means that cells \changed{can}{are
  allowed to} serve only a fraction of the traffic generated in the
area corresponding to some TP. In other words, we do not use a
all-or-nothing approach, where cells should serve either all
users or no users in the area corresponding to a TP.

\section{Load-Aware Energy-Efficiency Optimization}
\label{sec:sif}
The model presented in Sec.~\ref{sec:model} assumes the worst-case interference
in a fully loaded system, which leads to a lower bound on the link
spectral efficiency (c.f. Assumption \ref{as:worst_case_inter}). As
pointed out in Remark \ref{rem:worst_interference}, the main rationale
behind this approach is the need for avoiding coverage holes when
network elements are deactivated. The price is a sub-optimal
performance in terms of energy efficiency because the interference is
overestimated, and therefore users may use more resource blocks than
required to keep their minimum data rate requirements. An immediate
consequence of this is that more cells are activated than are
necessary for meeting the minimum rate requirements at the TPs.
In this section, we extend the optimization problem in
(\ref{eq.smm_prob}) to incorporate more precise estimates of the load
induced by a given user-cell assignment, which is not a
trivial task because it involves load computation (with fixed
assignments) that requires the solution of a system of nonlinear
equations \cite{Majewski2010,siomina12,CSSET14twct} (note that we can
easily estimate the link spectral efficiency from the load by using
\refeq{eq:specEffPL}).

In what follows, we propose an approach that typically yields good
approximations of the true link spectral efficiencies. The idea is to
use a two-step alternating iterative scheme:
\begin{enumerate}[\emph{Step} 1]
\item Compute the link spectral efficiency
  $\forall_{i\in\setm,j\in\setn}~\omega_{i,j}(\signal{\rho})$ defined
  in (\ref{eq:specEffPL}) for the load value obtained in the previous
  iteration of Step 2 of the algorithm (in the first iteration of the
  algorithm, we can use the worst-case spectral efficiency) and solve
  Problem (\ref{eq.smm_prob}) with these (fixed) link spectral
  efficiencies to obtain an TP-cell
  assignment $\ma{X}$. 
\item For the TP-cell assignment obtained in Step 1, compute
  the load induced by this assignment.
\end{enumerate}
Regarding the load computation in Step 2, we use the fact that the
load $\signal{\rho}$ induced by a given assignment $\ma{X}$ is a
fixed point of the following standard interference mapping (see
\cite{CSSET14twct,fettweis13} and the references therein for further
details):
\begin{equation*}
 \mathcal{J}:\R_+^M \to \R_{++}^M: \,
  \signal{\rho}\mapsto [I_1(\signal{\rho})~\ldots~{I}_M(\signal{\rho})]^T\,,
\end{equation*}
where
\begin{equation*}
I_i(\signal{\rho}) := \min\left\{
\sum_{j\in\setn} \frac{\lambda_{i,j}~x_{i,j}}{\log_2 \Bigl(1+\frac{1}{\eta_{i,j}^\mathrm{SINR}} {\frac{P_i~ g_{i,j}}{\sum_{k\in\setm\backslash\{i\}} P_k~ g_{k,j} ~ \rho_k + \sigma^2}}\Bigr)},\Gamma\right\}.
\end{equation*}
$\Gamma$ is a large constant and $\lambda_{i,j}:=\frac{r_{j}}{B_i ~ \eta_{i,j}^\mathrm{BW} }$. Since
$\mathcal{J}$ is a standard interference mapping and $I_i(\signal{\rho})$ 
is bounded above, we conclude that the fixed-point
always exists and is \emph{unique} \cite{yates95,martin11}. Moreover,
efficient iterative methods are known to approach the fixed point with
an arbitrary precision \cite{yates95,martin11}. We summarize the
heuristic proposed in this section in Alg.~\ref{alg:alternate}.

\begin{algorithm}
\caption{Load-aware energy minimization}
\begin{algorithmic}[1]
  \REQUIRE Worst-case spectral efficiency $\signal{\omega}^{(-1)} = \signal{\omega}(\signal{1}) 
	$. Maximum number of iterations $Z$.
  \ENSURE Network configuration $\ma{X}^{(Z)}$ with low energy
  consumption.  \FOR{$n=0:Z$} 
  \STATE Use $\signal{\omega}^{(n-1)}$ to construct Problem
  (\ref{eq:problem_original}).
  \STATE Use Alg.~\ref{alg:networkReconfiguration} to obtain
  $\ma{X}^{(n)}$ and remove deactivated cells from the set of
  cells to be considered in subsequent iterations.
  \STATE Compute the new link spectral efficiency
  $\signal{\omega}^{(n)}$ for the assignment $\ma{X}^{(n)}$ by
  computing the fixed point of the standard interference mapping
  $\mathcal{J}$.
\ENDFOR
\STATE Return the network configuration resulting from $\ma{X}^{(Z)}$.
\end{algorithmic}
\label{alg:alternate}
\end{algorithm}

\section{Numerical Evaluation}
\label{sec:numericalResults}
In the following we present a numerical evaluation of the performance of the proposed algorithm in different networks. We start by outlining the basic simulation scenario followed by a comparison with two reference schemes with respect to the energy savings and computational time. Next, we present the ability of the proposed algorithm to incorporate a variety of different base station energy consumption models. Finally, we show the performance gains achieved by applying 
Alg.~\ref{alg:alternate} from Sec.~\ref{sec:sif}.

\subsection{Basic Simulation Scenario}
The simulated network is located in a
square-shaped area of size 2km$\times$2km, where $L$ base stations are
placed at locations chosen uniformly at random. Unless stated otherwise,
each base station has three cells directed at $0^\circ$, $120^\circ$ and 
$240^\circ$ respectively. Traffic generated by
users is represented by $N$ TPs on an irregular grid. Hence,
each TP represents the traffic requirements of an area of
different size.  To obtain spatially varying traffic requirements, we
use the following traffic model in each run of the simulations. We
define three circular hot-spot areas with centers chosen uniformly at
random within the area. There are two types of TPs:
``hot-spot TPs (HTP)'' and ``standard TPs
(STP)''. Each TP in the simulation has probability 0.3 of
being a HTP and probability 0.7 of being a STP. While the position of
STP is chosen uniformly at random within the whole area, a HTP can be
assigned uniformly at random to one of three hot-spot area. Its final
position is determined in polar coordinates by sampling the distance
from the hot-spot center from a normal distribution and the angle from
a uniform distribution. We use a wrap around model to avoid boundary
effects and determine the location of TPs to be placed outside
the square-shaped area.  The data rate requirements of TPs are
derived from a normal distribution with $\mu_{\text{d}} =
128~\text{kbps}$ and variance $\sigma_{\text{d}}^2= 32 ~\text{kbps}^2$
with a lower bound of $1~\text{kbps}$. The signal attenuation for
links between cells and TPs follows the ITU
propagation model for urban macro cell environments with a 
horizontal antenna pattern for 3-sector cell sites with fixed antenna patterns \cite{3GPP2010}.

Unless otherwise stated, we use the following simulation parameters:
$\epsilon^\star = 10^{-3}$, $\epsilon = 10^{-3}$, $B_i =
20\text{MHz}$, $P_i = 40dB$, $\eta_{\text{SINR}}= 1$,
$\eta_{\text{BW}}= 0.83$, $c_i = 500\text{W}$ and $e_i = 280\text{W}$. The values of the
last six parameters have been chosen to mimic the behavior of
commercial LTE systems. Furthermore, we use
$f_i(\rho_i)=564~\rho_i$ to model the load-dependent energy
consumption, which is a value similar to the dynamic energy
consumption of current macro cells with 6 transmit antennas \cite{earthd23}.

The proposed algorithms are compared with a solution of the original 
problem in \refeq{eq:problem_original} and, where possible, with the centralized cell zooming 
approach from \cite{Niu2010}. The solution to the problem
in \refeq{eq:problem_original} is obtained by using
Matlab 2013a in combination with IBM's CPLEX on a Intel Core i7 PC
with four cores. As shown later in this section, the computational
time to solve \refeq{eq:problem_original} grows fast with the problem
size. Therefore, to solve the problem in \refeq{eq:problem_original}
in a reasonable time for comparison purposes, we confine our attention
to small networks with $M=102$ cells ($L=34$ base stations) and $N=100$ TPs,
unless otherwise stated. We obtained the 95\% confidence intervals depicted
in the figures by applying the bias corrected and accelerated
bootstrap method \cite{efron1987} to the outcome of 100 independent
runs of the simulations. Results related to the overall network energy
consumption will be normalized to the energy consumption of the
network when all cells are active and fully loaded.
\begin{definition}[Normalized network energy consumption]
\label{def:normalized_energy}
Given a TP assignment $\ma{X}$ inducing cell load $\signal{\rho}$ and given the resulting network energy consumption
$E(\signal{\rho})$, the normalized network energy consumption is defined to be
\begin{equation*}
E_{\text{norm}}(\signal{\rho}) := \frac{E(\signal{\rho})}{E(\signal{1})} = \frac{E(\signal{\rho})}{\sum_{l \in \setl} c_l + \sum_{i \in \setm}  \bigl(e_i + f_{i}(\signal{1})\bigr)},
\end{equation*}
where the term in the denominator is the energy consumption for a fully loaded
system ($\signal{\rho} = \signal{1}$).
\end{definition}

We refer to the sparsity supporting
majorization-minimization algorithm as \emph{``sMM''} and to any
algorithm that solves \refeq{eq:problem_original} directly as \emph{``MIP''}
algorithm (MIP: mixed-integer programming).  We refer to solutions obtained
by the centralized cell zooming algorithm in \cite{Niu2010} as \emph{``cCZ''}. The alternating approach
proposed in Sec.~\ref{sec:sif} is referred to as \emph{``alternating sMM''}
algorithm.

\subsection{Computational performance comparison between \textit{sMM}, \textit{cCZ} and \textit{MIP}}
\label{sec:numRes_quality}

The \textit{cCZ} has limited capability to incorporate different energy consumption models and base
stations with several sectors, so we confine ourselves to a simple base station model. 
We assume a homogeneous network model under which all base stations have only one omni-directional cell, and all base
stations have the same energy consumption model. More precise, we use $|\setl| = M =100$, $|\sets_l| = 1$ and \refeq{eq:totalNetEnergy} with $c_l = 500$, $e_i = 280$, $f_i(\rho_i) = 0$ ($l\in\setl$, $i\in\setm$).\par
\begin{figure}[tb]
	\centering
	\includegraphics[width=0.5\columnwidth]{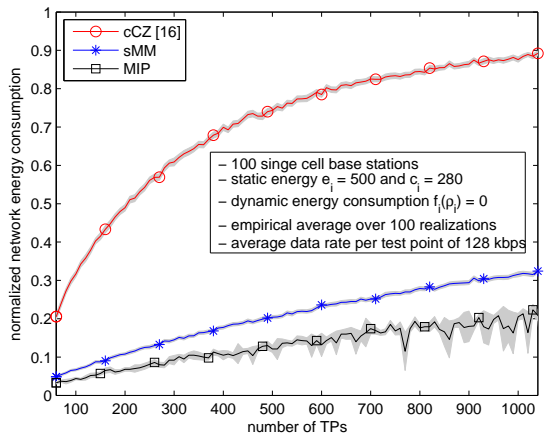}
	\caption{Comparison of normalized network energy consumption obtained with
          the \textit{sMM} algorithm, the \textit{cCZ} algorithm and the solution of the
          \textit{MIP} problem for increasing number of TPs.
          Normalization with respect to the network energy
          consumption for a fully loaded system ($\signal{\rho} =
          \signal{1}$) when all cells are active. }
	\label{fig:ee_savings}
\end{figure}

To show trends, we start with the standard setup described above, and we gradually increase the number of TPs in the system. 
Fig.~\ref{fig:ee_savings} shows the \emph{normalized network energy consumption}.
As expected, the normalized network energy consumption for all three algorithms
increase as the number of TPs increases. This is intuitive
because additional TPs add extra rate requirements that increase
the total system load, which in turn reduces the redundancy in the
network to be exploited for energy savings. The proposed \textit{sMM}
algorithm as well as the \textit{MIP} algorithm provide network configurations that exhibit much smaller
normalized network energy consumption when compared with the network
configurations obtained with the \textit{cCZ} algorithm. The smallest energy consumptions are achieved with the \textit{MIP} algorithm, which outperforms the proposed \textit{sMM} algorithm. For the scenario with 200 TPs,
the \textit{sMM} algorithm results in normalized network energy
consumption of 12\% on average. For the same number of TPs, the
average normalized energy consumption under the \textit{cCZ} and \textit{MIP} algorithm are 49\%
and 7\%, respectively. Similarly, for 1000 TPs, the resulting average normalized network energy consumption
of 31\% for the \textit{sMM} algorithm is still larger than the
21\% normalized energy consumption corresponding to the
\textit{MIP} solutions. However, it is still much smaller than \textit{cCZ} with 88\% normalized energy consumption. 
These results emphasize that the \textit{sMM} algorithm is a suboptimal heuristic, which is able to find network configurations
consuming low energy. Even though the resulting network energy consumption is not globally optimal, it
shows much larger energy savings than the comparison scheme \textit{cCZ}.

\begin{figure}[tb]
	\centering
	\includegraphics[width=0.5\columnwidth]{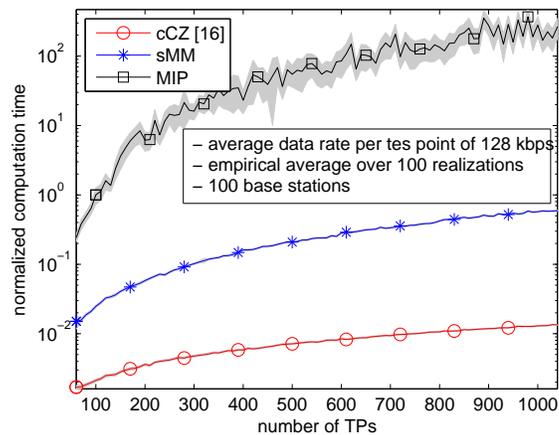}
	\caption{Comparison of normalized computation time to obtain
          results with the \textit{sMM} algorithm, the \textit{cCz} algorithm and the direct solution of the \textit{MIP} problem. Normalization with
          respect to the empirical average of the \textit{MIP}'s computation time 
					for 100 cells and 100 TPs over 100 realizations .}
	\label{fig:compTime}
\end{figure}

The main advantage of the proposed \textit{sMM} algorithm is its fairly low
computational complexity, which is directly affecting the time
required to obtain an optimization result.  Fig.~\ref{fig:compTime}
depicts the normalized time needed to obtain the results
of Fig.~\ref{fig:ee_savings}. This time is
normalized with respect to the computation time of the \textit{MIP}
algorithm with 100 cells and 100 TPs.  The
\textit{sMM} algorithm always provides results in a substantially
shorter time than the \textit{MIP} algorithm.  Even for a relatively
small scenario of 100 cells and 300 TPs, the
computation time is already about 200 times larger for the
\textit{MIP} algorithm compared to the proposed \textit{sMM}
algorithm.  For larger setups with 1000 TPs the normalized
time to solve the \textit{MIP} was $\approx 237$ compared to $\approx
0.49$ for the \textit{sMM} algorithm, which is an approximate 488 fold
reduction in the computation time. We emphasize that the simulated
scenarios are small and the computation of the MIP solution becomes
infeasible in practical scenarios. Already for a network with 200 cells
and 10,000 TPs, the sMM algorithm provided a solution
in about 13s, whereas the MIP algorithm could not find a solution
within one hour.
Compared to the \textit{cCZ} algorithm the proposed \textit{sMM} algorithm takes longer time due to the lower complexity heuristic used in the \textit{cCZ} algorithm. For a scenario of 300 TPs the average computation time is about 22 times larger for the \textit{sMM} algorithm and with 1000 TPs it is about 43 times larger. 
However, with typical values of less than 1s, the computation time is still reasonably small to allow for an online implementation. Considering the advantages in energy savings, as seen from Fig.~\ref{fig:ee_savings}, the proposed \textit{sMM} algorithm presents a good trade off between computation time and energy savings.

\subsection{Cells with different sources of energy consumption}
\label{sec:numRes_energy}
In contrast to other approaches to the problem of energy-efficient
network topology control, our optimization framework can easily deal
with heterogeneous networks in which cells have different
static and load-dependent energy consumptions in \refeq{eq:totalNetEnergy}.
In other words, the
proposed \textit{sMM} algorithm can cope with different energy
consumption models of cells. It can select those network
configurations that exhibit as low overall energy consumption as
possible.  To illustrate the impact of different energy consumption
models on the optimization result, we start by varying the static energy
consumption of all cells, while keeping the
load-dependent energy consumption fixed. 
Later
in this section, we show the impact of the load-dependent energy
consumption by changing the weight of the load dependent part relative
to the static part.

\begin{figure}[tb]
	\centering
	\includegraphics[width=0.7\columnwidth]{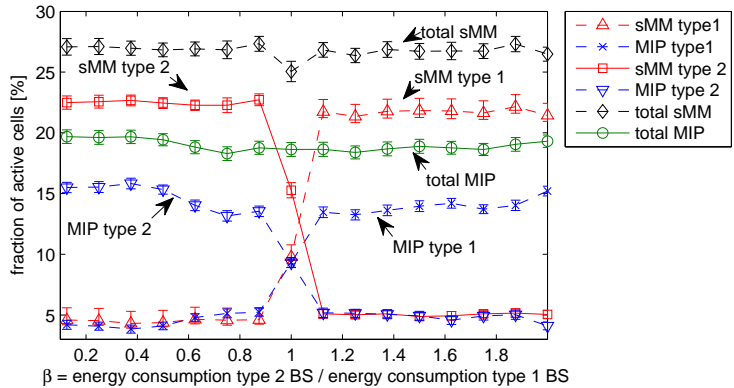}
	\caption{Fraction of active \textit{type 1} and \textit{type
            2} cells in the final solution obtained with sMM
          and MIP. Deployment uniformly at random for \textit{type 1}
          and \textit{type 2} cells.}
	\label{fig:differentenergy}
\end{figure}

To study the impact of the static
energy consumption of cells $e_i$, in the
following simulations we use single cell omnidirectional 
base stations, and we set the 
load-dependent part for all cells and the common static
part at base stations to zero $f(\rho_i)=0$ and $c_l=0$.  The static energy
consumption of half of
the cells is varied, while the static energy consumption of
the other half remains unchanged. We refer to the cells with standard
fixed energy consumption as \textit{type 1}, while \textit{type 2} is
used to refer to cells with a varying energy consumption. The
energy consumption of \textit{type 2} cells is specified
relative to that of \textit{type 1} cells. More precisely,
an energy consumption relation of $\beta =0.5$ means that if
$c_i=780$W for \textit{type 1} cells, then $c_i=390$W for
\textit{type 2} cells. The results for a scenario consisting
of 100 cells and 100 TPs are shown in
Fig.~\ref{fig:differentenergy}.
The simulation confirms the ability of our optimization framework to
incorporate different static energy consumptions. When all cells
consume the same amount of energy ($\beta=1$), the algorithm
makes no difference between \textit{type 1} and \textit{type 2} cells.
The energy consumption of \textit{type 1} and \textit{type 2} cells
is roughly the same indicating that equally many \textit{type 1} and \textit{type 2} cells
are active in the obtained solution.
In contrast, if \textit{type 2} cells consume less energy than
\textit{type 1} ($\beta < 1$), then the algorithm prefers to
deactivate \textit{type 1} cells, while attempting to keep
\textit{type 2} cells active. Obviously, if $\beta>1$, the
situation is reversed in the sense that, if possible, \textit{type 2}
cells are preferably selected for deactivation. \par
\begin{figure}[tb]
	\centering
	\includegraphics[width=0.5\columnwidth]{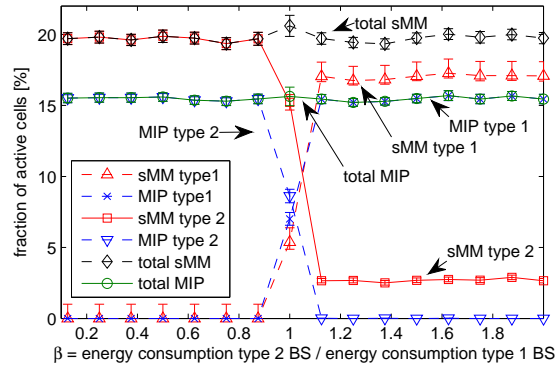}
	\caption{Fraction of \textit{type 1} and \textit{type 2} cells
          in the final solution obtained with sMM and
          MIP. Deployment of \textit{type 1} cells uniformly
          at random and \textit{type 2} cells are co-located
          with \textit{type 1}.  }
	\label{fig:differentenergy2}
\end{figure}

The differentiation becomes even more evident for cell
deployments, where \textit{type 1} and \textit{type 2} cells
are co-located. In such a case, two cells of different type
are located at the same site and are ``exchangeable'' with respect to
the service provided to the TPs (recall that we use omnidirectional cells in these simulations). In other words, if a TP
is assigned to a location with two co-located cells,
then it does not matter which cell is used to provide the
service to the TP. This implies that the decision whether to
deactivate a cell or not should depend only on the energy
consumption of this cell in relation to its co-located cell\footnote{
Even though such setups are unlikely in practice, we use it for reasons of 
illustration.}.
The simulations with such a deployment are shown in
Fig.~\ref{fig:differentenergy2}, where we see that, for $\beta<1$,
there is no active cell of \textit{type 1}, while,
for $\beta>1$, \textit{type 2} cells consume more energy and
the simulations confirm that the algorithms clearly prefer to activate
\textit{type 1} cell.

\begin{figure}[tb]
	\centering
		\includegraphics[width=0.5\columnwidth]{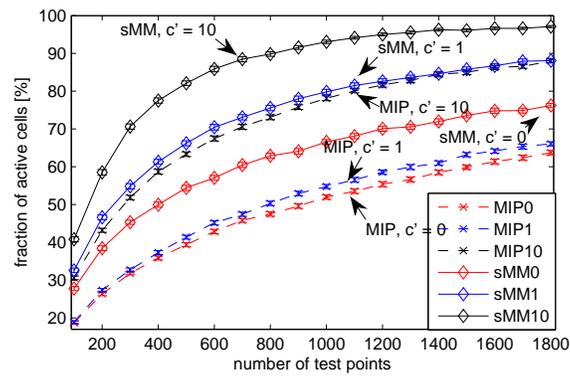}
                \caption{Fraction of active cells for
                  different dynamic energy consumption c' with
                  increasing number of TPs.}
	\label{fig:bs_inc_dynamic_var}
\end{figure}

\begin{figure}[tb]
	\centering
		\includegraphics[width=0.5\columnwidth]{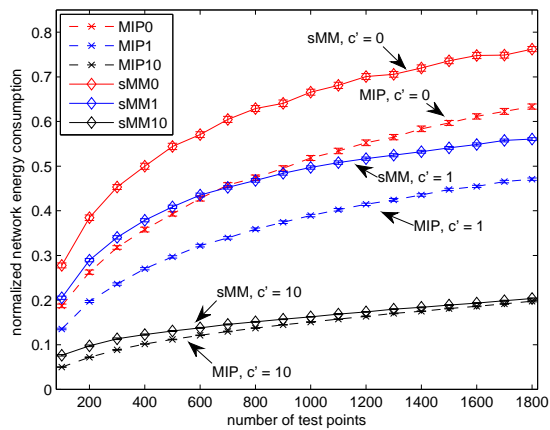}
                \caption{Normalized network energy consumption for
                  different dynamic energy consumption weights c' with
                  increasing number of TPs. Normalization with
                  respect to the energy consumption when all cells
                  are active.}
	\label{fig:bs_inc_dynamic_eng}
\end{figure}

To obtain insight into the impact of the load-dependent energy
consumption, we fix the static energy consumption of a single-cell 
omnidirectional base station to be $e_i = 780$W and 
$c_l = 0$W,
and we vary the load-dependent energy consumption
$f_i(\signal{\rho})=564\,c'\,\rho_i$ by letting $c'$ take values on $c' \in
\{ 0, 1, 10\}$. For an increasing number
of TPs, Fig.\ref{fig:bs_inc_dynamic_var} shows the fraction of
active cells, while the normalized network energy consumption is shown in
Fig.\ref{fig:bs_inc_dynamic_eng}.
First we observe that the network energy consumption always
increases with an increasing number of TPs, which is in fact
no surprise.  Moreover, the fraction of active cell increases
with $c'$ for both the \textit{sMM} algorithm and the \textit{MIP}
algorithm. An examination of the objective function in \ref{eq:p1.0}
shows that this is what we expect because if the ratio of the
load-dependent energy consumption becomes larger relative to the
static one, then the algorithm tends to increase the fraction of active
cells for an improved load balancing in order to keep the load
of each active cell at a relatively low level. In other words,
instead of deactivating as many cells as possible to minimize
the static energy consumption, the algorithm deactivates the cells
to find the best possible balance between the static and
load-dependent energy consumption. This can be observed in
Fig.\ref{fig:bs_inc_dynamic_var}, where we can see that the higher is
the load-dependent energy consumption (which is reflected by $c'\geq
0$), the more cells are activated under both the \textit{sMM}
algorithm and \textit{MIP} algorithm. In particular, if $c'=10$,
then the fraction of active cells is significantly increased
compared with the situation, in which the load-dependent energy
consumption is negligible ($c'=0$).

\subsection{Alternating \textit{sMM} algorithm}
\label{sec:numRes_loadrecompp}
We now study the performance of the alternating sMM algorithm
presented in Sec.~\ref{sec:sif}. The standard simulation parameters
are used with a total number of $Z=10$ iterations. To show the effect
of different TP requirements, we performed simulations under our standard simulation setup for
different mean data rates $\mu_d$ at TPs and, for each mean data rate,
we used 100 different realizations of the simulation scenario. The
initial link spectral efficiency is computed based on the worst-case
interference according to Eq.~(\ref{eq:specEffPL}). Our goal is to
show the huge potential for energy savings when the actual load is
estimated as in Alg.~\ref{alg:alternate}, instead of assuming the
worst-case interference scenario, which corresponds to the full-loaded
system (see Definition \ref{as:worst_case_inter}).
\begin{figure}[tb]
	\centering
	\includegraphics[width=0.55\columnwidth]{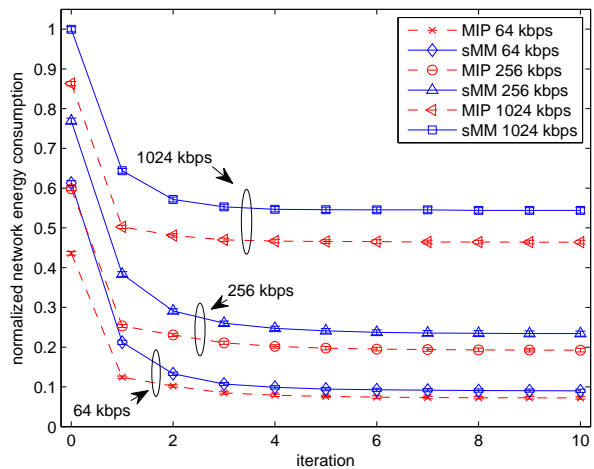}
	\caption{Normalized network energy consumption when applying
          Alg.~\ref{alg:alternate} when using the sMM algorithm
          compared to the using the MIP solution in each iteration.
          Normalization with respect to the energy consumption when
          all cells are active.}
	\label{fig:loadrecomp}
\end{figure}
The outcome of the simulation is depicted in
Fig.~\ref{fig:loadrecomp}, which includes the 95\% confidence
level and shows the normalized network energy consumption with respect
to the energy consumption when all cells are active. We can see that the application of the techniques from
Sec.~\ref{sec:sif} leads to a significant reduction of the normalized
network energy consumption for both the sMM algorithm and the optimal
MIP solution. Furthermore, the largest reduction was always observed
after the first
iteration, 
which shows that the worst-case interference assumption is very
conservative and the load estimation may lead to considerable
performance gains.

\section{Concluding Remarks}
\label{sec:conclusion}
We have introduced an optimization framework for enhancing the energy
efficiency of cellular networks. In wireless systems, problems of this
type are hard to solve because of their combinatorial nature and the
nontrivial interference coupling among cells. Indeed, even
with a simplifying assumption of the worst-case interference, the
energy saving problem is a mixed integer programming problem that is
strongly related to the bin-packing problem, which in turn is known to
be NP-hard. As a result, we cannot expect to find optimal
solutions quickly, we focused in this study on fast sub-optimal
heuristics. Unlike many existing approaches in the literature, the
proposed methods can naturally consider both the dynamic and static
energy consumption of base stations with multiple cells in heterogeneous networks.

In the first proposed heuristic, we relaxed the mixed integer
programming problem to a form suitable for the application of
majorization-minimization techniques. The resulting algorithm requires
the solution of a series of linear programming problems that can be
efficiently solved with standard mathematical solvers. Therefore, it
can be applied to large-scale problems, and it is also suitable for
online operation. One limitation of this first method is that it uses
the worst-case interference scenario, so it can be too conservative in
terms of energy savings. To address this limitation, we also proposed
a two-step alternating approach that obtain accurate values of the
spectral efficiency of links by using the framework of standard
interference functions. Simulations show that the proposed fast
heuristics are able to obtain network configurations that are
competitive in terms of energy consumption against optimal
algorithms. \par

\section*{\normalsize Acknowledgments}
This work has been partly supported by the framework of the research
project ComGreen under the grant-number 01ME11010, which is funded by
the German Federal Ministry of Economics and Technology (BMWi).
Part of this work has been performed in the framework of the FP7 project 
ICT-317669 METIS, which is partly funded by the European Union. The authors would 
like to acknowledge the contributions of their colleagues in METIS, although the 
views expressed are those of the authors and do not necessarily represent the 
project.
\ifCLASSOPTIONcaptionsoff
  \newpage
\fi

\appendix
\label{sec:MMalgo}
Here we briefly summarize the majorization-minimization (MM) algorithm
\cite{Hunter2004}, which can be seen as a generalization of the well-
known expectation-maximization (EM) algorithm. The presentation that
follows is heavily based on that in the study in \cite{sriperumbudur}
(see also \cite{Pollakis2012,CSSET14twct}).\par

Suppose that the objective is to minimize a \quest{Do we need
  continuity?} function $h:\mathcal{X}\rightarrow \R$, where
$\mathcal{X}\subset\R^N$. Assume that there exists a solution to this
optimization problem, and let $\ve{x}^\star\in \mathcal{X}$ be a
global minimizer of $h$; i.e., $h(\ve{x}^\star)\leq h(\ve{x})$ for
every $\ve{x}\in\mathcal{X}$. Unless $h$ has a special structure that
can be exploited (e.g. convexity), finding $\ve{x}^\star$ is
computationally intractable in general \cite{Rockafellar1970}. Hence,
we typically have to content ourselves with generating a sequence of
vectors with non-increasing objective
value. 
To this end, we can use the majorization-minimization (MM) technique,
which drives $h$ downhill 
with the help of a majorizing function $g:\mathcal{X}
\times\mathcal{X}\rightarrow \R$. In more detail, we say that $g$ is majorizing function for $h$ if it satisfies the following properties:
\begin{enumerate}[C.1]
\item $g$ majorizes $h$ at every point in $\mathcal{X}$, i.e.
\begin{align}
 \label{eq:mmp1}
 h(\ve{x})\le g(\ve{x},\ve{y}),\quad\forall\ve{x},\ve{y}\in \mathcal{X},
\end{align}
\item $g$ and $h$ coincide at $(\ve{x},\ve{x})$ so that
\begin{align}
 \label{eq:mmp2}
 h(\ve{x})=g(\ve{x},\ve{x}),\quad\forall\ve{x}\in\mathcal{X}.
\end{align}
\end{enumerate}
By starting from a feasible point $\ve{x}^{(0)}\in\mathcal{X}$, the MM
algorithm generates a sequence
$\left\{\ve{x}^{(n)}\right\}_{n\in\Natural}\subset\mathcal{X}$ with
monotone decreasing function values $h(\ve{x}^{(n)})$ according to (we assume that the optimization problems have a solution)
\begin{align}
\label{eq:mmit}
 \ve{x}^{(n+1)} \in \arg\min_{\ve{x}\in\mathcal{X}}g(\ve{x},\ve{x}^{(n)})\,.
\end{align}
Irrespective of the choice of $g$, we can easily verify monotonicity
of the objective value with the help of (\ref{eq:mmp1}),
(\ref{eq:mmp2}) and (\ref{eq:mmit}): $h(\ve{x}^{(n)})=g(\ve{x}^{(n)},\ve{x}^{(n)})\geq g(\ve{x}^{(n+1)}, \ve{x}^{(n)}) 
\geq g(\ve{x}^{(n+1)},\ve{x}^{(n+1)})  = h(\ve{x}^{(n+1)}) $.
Therefore, since the function $h$ is bounded below when restricted to
$\mathcal{X}$ by assumption, we can conclude that $h(\ve{x}^{(n)})\to
c\in\R$ for some $c\ge h(\ve{x}^\star)$ as $n\to\infty$. However, we
emphasize that this in general does not imply the convergence of the
sequence $\left\{\ve{x}^{(n)}\right\}$.
\par 
The choice of the function $g$ is problem dependent, but it should be
sufficiently structured in order to make the optimization problem in
\refeq{eq:mmit} tractable. In particular, in our study we  deal with concave and
continuously differentiable functions $h$. In such cases, a natural
choice for $g$ satisfying (\ref{eq:mmp1}) and (\ref{eq:mmp2}) is
\begin{align}
\label{eq.g_concave}
 g(\ve{x},\ve{y}) = h(\ve{y})+\nabla h(\ve{y})^T(\ve{x}-\ve{y}).
\end{align}
This particular choice is common in, for example, sparse signal
recovery \cite{Candes2008}.  \par
\begin{remark}
\label{re:mm_valid_dec}
We note that, instead of solving the optimization problem in
\refeq{eq:mmit} exactly, {it is sufficient for the monotonicity of the
  sequence $\{h(\signal{x}^{(n)})\}$} that
$g(\ve{x}^{(n+1)},\ve{x}^{(n)})\leq g(\ve{x}^{(n)},\ve{x}^{(n)})$ for
every $n\in\NN$. This observation is relevant if the right-hand side
of \refeq{eq:mmit} can only be solved asymptotically, in which case
the iteration can be truncated whenever the above inequality is
satisfied.
\end{remark}

\bibliographystyle{IEEEtran}
\bibliography{IEEEabrv,references}

\end{document}